# QoS Guaranteed Intelligent Routing Using Hybrid PSO-GA in Wireless Mesh Networks

*V. Sarasvathi[1], N. Ch. S. N. Iyengar[2], Snehanshu Saha[1]*

[1] *Department of Computer Science and Engineering, PESIT Bangalore South Campus, Bangalore-560 100, India.*
[2] *School of Computing Science & Engineering, VIT Unversity, Vellore-632014, Tamilnadu , India*
*Emails: sarasvathiram@gmail.com, nchsniyengar48@gmail.com, snehanshusaha@pes.edu*

**Abstract:** *In Multi-Channel Multi-Radio Wireless Mesh Networks (MCMR-WMN), finding the optimal routing by satisfying the Quality of Service (QoS) constraints is an ambitious task. Multiple paths are available from the source node to the gateway for reliability, and sometimes it is necessary to deal with failures of the link in WMN. A major challenge in a MCMR-WMN is finding the routing with QoS satisfied and an interference free path from the redundant paths, in order to transmit the packets through this path. The Particle Swarm Optimization (PSO) is an optimization technique to find the candidate solution in the search space optimally, and it applies artificial intelligence to solve the routing problem. On the other hand, the Genetic Algorithm (GA) is a population based meta-heuristic optimization algorithm inspired by the natural evolution, such as selection, mutation and crossover. PSO can easily fall into a local optimal solution, at the same time GA is not suitable for dynamic data due to the underlying dynamic network. In this paper we propose an optimal intelligent routing, using a Hybrid PSO-GA, which also meets the QoS constraints. Moreover, it integrates the strength of PSO and GA. The QoS constraints, such as bandwidth, delay, jitter and interference are transformed into penalty functions. The simulation results show that the hybrid approach outperforms PSO and GA individually, and it takes less convergence time comparatively, keeping away from converging prematurely.*

**Keywords:** *Wireless mesh networks, Multi-radio, Multi-channel, Particle swarm optimization, Genetic algorithm, Quality of service.*

## 1. Introduction

The next generation technology for Internet service provisioning focuses on every common man to get the access facility with every gadget he possesses. Even though the count of Internet Service Providers grows on the market, the traditional technology is still restricted by the bandwidth and the cost constraints. The 2G GPRS cellular system provides data rates up to 114 kbps and the 3G networks provide data rate ranges from 384 kbps up to 2 Mbps. The spotted drawback in 2G and 3G services is the bundling of multiple services into the available bandwidth in a reduced timeslot. These types of services are suitable only for web browsing and email sending, but not for gaming and video streaming applications. For example, viewing You Tube videos continuously for 10 minutes consumes from 30 up to 35 MB approximately and it is evident that the cable Internet access, such as cable broadband, DSL and Wi-Fi are desirable for high speed Internet services.

Installing fiber Optic Cables, building Cellular Base Stations and maintaining the wired line for service providing drastically increases the expense of the infrastructure itself. The innovative and cutting edge technology in the field is the advent of Wireless Mesh Network (WMN) [1] which offers high bandwidth and imbibes low cost of deployment and maintenance. WMN is the next generation network that aims at providing high speed Internet access to any user and features self-configuring and self-healing properties. The major challenge faced by the researchers in the development of WMNs was to select an optimal path which avoids interference and also increases the performance. The interference is not only between the neighbouring links which are assigned to the same channel, but also from the adjacent channels and self- interference. An optimal method is needed for routing and the algorithms must also be fast enough to converge for large WMN.

The WMN concerns about the capacity, which is achieved through the multi-channel and multi-radio, and the characteristics that affect the capacity of the network are the bandwidth, interference, delay and jitter. The WMN is rapidly emerging in recent years due to its potential applications, such as community networks, broadband home networks and commercial networks. Since these networks require low investment and minimal infrastructure for deployment, they can be used to leverage the converting cities into smart cities. It acts as an interface between the users who want to connect their laptops and smart phones to access applications over Internet. The biggest challenge in designing a smart city is achieving the optimal routing capability with efficient resource utilization and at the same time the Quality of Service (QoS) also being fulfilled.

WMNs comprise of two types of nodes: mesh routers and mesh clients. The mesh routers perform routing functions to support mesh networking and they are relatively static nodes. They primarily act as the mesh backbone for mobile clients and in turn are responsible for establishing the mesh connectivity among the clients. The route selection is based on link discovery and connected with the fact that the link which does not interfere with other transmissions would provide a higher throughput. Simultaneous data transmission and reception is possible by routers because they are basically built-in with multiple radios and abundant power. Mesh clients can play the dual role, for they can act as a host and also as a router.

In this paper an example of IEEE802.11b/g wireless technologies has been considered, which operates on 2.4 GHz spectrum frequency, and the spectrum band was split into 11 channels, out of which 3 channels are non-overlapping. Because of the limited availability of non-overlapping channels, all the nodes were assigned to the same channel and that lead to more performance degradation of the networks. When the same channel was assigned to all the interfaces, the performance of the router was more degraded than with the partially overlapping channel assignment. When MCMR-WMN considered along with the partially overlapping channels assigned to the radio on a mesh router, it was observed that the low cost IEEE802.11b/g mesh networking hardware greatly improves the capacity of the infrastructure mesh networks compared to other existing technologies.

When the routers are equipped with multiple radios, the traditional shortest path routing algorithm does not work well and the traditional routing algorithm finds a path without considering the different channels assigned to the radios. In this paper we propose an optimal intelligent routing algorithm using PSO-GA which finds the most optimal routes and satisfies QoS constraints.

The remainder of the paper is organized as follows. In Section 2 some of the meta-heuristic algorithms used in routing problems are presented. Section 3 elaborates the existing routing algorithms. Sections 4 and 5 describe the original PSO algorithm and GA algorithm respectively. Section 6 presents the Hybrid PSO-GA algorithm. In Section 7 the proposed QoS intelligent routing, the determination of fitness functions, finding of successive particles using a $\oplus$ operator and a crossover operator are presented. Section 8 elaborates the simulation and the results. The conclusion of the paper is found in Section 9.

## 2. Related work

There is a quite large number of meta-heuristics methods available for optimization of the routing algorithms. They fall into different categories as follows: Simulated annealing, Evolutionary algorithms, Tabu search and Swarm intelligence algorithms.

Simulated annealing is a meta-heuristic method with a probability feature to find a good optimized solution in a large search space. It is an adaptable method for networks and more suitable for the discrete type; since the nodes in a network are discrete, the simulated annealing method can be applied here. The simulated annealing algorithm is more appropriate for time-bound solutions, it may not guarantee the best solution, but it gives a better solution within a given time limit.

In [2] the link instability and link quality were considered for the routing algorithm in WMN, and Ant Colony Optimization (ACO), and Simulated Annealing (SA) algorithms were presented. SA was integrated with ACO to speed up the convergence rate.

The improved Dijkstra's algorithm was used in [3] to find the shortest path from the end node to the gateway which improved the reliability of data transmission, and also dealt efficiently with the link failures. Here, the ACO was combined with an enhanced Dijkstra's algorithm, the route setup process was carried out by an enhanced Dijkstra's algorithm and the route exploration and maintenance was performed by ACO. Initially, the network manager identified the shortest path to each end node using the enhanced Dijkstra's algorithm. After that, the ACO algorithm performed the searching for the remaining routes in the route exploration stage, which was simultaneously processed along with the data transmission. The topological changes in the network were considered in the route maintenance scheme and the response was efficient in a timely manner. There were two types of pheromones used in this algorithm: The regular pheromone and the virtual pheromone. The goodness of the routes was estimated by the regular pheromone and it also helped finding the route which can be used for data transmission. A virtual pheromone was presented to send the data which helped the ants in sampling the possible paths in networks.

B o k h a r i et al. [4] have proposed a distributed routing algorithm for finding the high throughput, less interference and load balanced path in WMN. This AntMesh proposal is based on the stochastic approach, which simultaneously performs the routing and data forwarding in a dynamic network. The ants utilize multiple channels efficiently with multiple interfaces in WMN. There are three types of ants used in this algorithm: Forward Smart Ants (FSA), Backward Smart Ants (BSA) and Hello Smart Ants (HSA). FSA travel from the source to the destination to determine the paths, BSA travel from the destination to the source for updating the routing tables and HSA collect the local link quality information to populate the link estimation table.

L i a n g   D a i et al. [5] have considered the dynamic nature of the traffic demand in optimal routing. It is based on two of the components in the framework: traffic estimation and routing optimization. The traffic estimation uses the time series analysis to predict the future traffic demand based on the historical data. It predicts the mean demand for a long period of time in dynamic networks and also statistical distribution. Two of the routing algorithms were presented in two different forms of the traffic demand estimation. First, the mean value of the traffic prediction was given as an input to the routing algorithm which was formulated as a linear programming problem to increase the data flow. Second, the traffic demand was characterized based on a random variable to incorporate the statistical distribution in the formation of the problem. Routing optimization balanced the traffic and ensures that minimum congestion will occur.

R a s t i n   P r i e s et al. [6] focused on routing and channel assignment in WMN to achieve the max-min fair throughput allocation. They have used the Genetic Approach to optimize the deployment of WMN. The Progressive filling algorithm was used to achieve the Max-min fairness; eight different fitness functions were used for optimizing the path from the source to the gateway – for example, the minimum, mean, and maximum throughput. Instead of 2-Point Crossover, two new cross variants, called Cell and Subtree Crossover were introduced which produce the best solution. When a large number of users per a gateway was used, the Sub tree Crossover method was chosen for better performance and the Cell Crossover gave the best solutions in scenarios with a smaller number of end-users per a gateway.

In most of the work above mentioned, the authors focused on routing in WMN with orthogonal channels assigned to the interfaces and the physical distance between the source and the gateway was considered without considering the interference on the link. The capacity of the link is not only based on the physical distance, but also the interference on the link. In contrast to the existing work, we focus on routing in MCMR-WMN with POC assigned; the radios are assigned with channels in such a way that it does not interfere with the neighbouring link.

We considered MCMR-WMN with partially overlapping channels assigned to the interfaces, based on the graph edge colouring method mentioned in our previous work [7]. The adjacent channel and self-interference is observed more when the partially overlapping channel is assigned to a radio. The shortest path may not be the interference free path, and it leads to a collision domain, so we propose a hybrid PSO-GA algorithm which finds a less interferenced path efficiently and also satisfies the bandwidth, delay and jitter request specified by the user.

## 3. Routing algorithms

Routing algorithms can be classified into three types, namely: a reactive algorithm, a proactive algorithm and a hybrid algorithm. The Proactive algorithms try to find the routes to all the other nodes, maintain more than one routing table and the routing information is updated at predefined intervals. The Reactive algorithms are on-demand based and they find the route only when it is needed. The AODV algorithms are based on-demand, which does not provide any QoS guarantee.

The existing routing algorithm is based on the statistical characteristic of a link and the interference is a key factor which influences the packet delivery between the mesh routers. Routing is challenging due to interference and the unpredictable nature of the wireless environment.

Routing in WMN is a multi-objective optimization problem, when it comes to satisfying the QoS constraints, such as bandwidth, delay, interference and jitter. Various kinds of meta-heuristic and natural mechanisms, such as Ant colony optimization, Particle swarm optimization, Genetic algorithms, and neural networks are used to find a good solution for a complex problem with fast convergence time. But these algorithms fail to solve the case of routing in MRMC-WMN with multiple objective functions subject to multiple QoS constraints. We propose a multi-objective optimization problem which minimizes the cost while maximizing the channel utilization, and also maximizes the network performance while minimizing the interference.

## 4. Particle swarm optimization

Particle Swarm Optimization (PSO) [8] was developed by Dr. Eberhart and Dr. Kennedy in 1995, motivated by the social behaviour of the species, such as bird flocking or fish schooling, used to solve the meta-heuristic optimization problem. PSO is an iterative process which guides to explore and exploit the search space. The group of entities in PSO are called particles which have a position and velocity, and each of the particles explores the solution in the multidimensional search by adjusting the position and velocity. The particle position gives a candidate solution in the search space, and the individual particles have no intelligence and it just follows the simple basic rules in a decentralized manner and acts based on the local information.

Each particle has memory and the previous state is remembered, the individuality retains the particle's previous best position and the sociality retains the neighbour's previous best position. Each of the particles remembers its best value using its own experience; the best value is represented by pbest and the position is represented by pbest$X$[], pbest$Y$[]. Each particle knows the global best position and it is represented by gbest. The gbest is the knowledge of the group and this knowledge is informed to all the individuals. PSO is the fastest search method for many complicated problems, and the performance of each particle is evaluated based on the fitness functions. At each iteration the particle's velocity and the position are updated using the formulas

(1) $$v_i(t+1) = \omega v_i(t) + c_1 r_1 [\, p_i(t) - x_i(t)\,] + c_2 r_2 [\, p_g(t) - x_i(t)\,],$$

(2) $$x_i(t+1) = x_i(t) + v_i(t+1),$$

where: $i = 1, 2, \ldots, N$; $t = 1, 2, 3, \ldots, T$; $N$ is the size of the swarm and $T$ is the limit of iteration; $p_i$ and $g_i$ are the local best and global best solutions; $c_1$ and $c_2$ are cognitive and social factors in the acceleration and these values are between 0 and 2; $r_1$ and $r_2$ represent two random numbers between 0 and 1; $w$ represents the inertia weight which balances the PSO algorithm between the local and global search. The largest value of the inertia weight leads to the global search and the smallest value facilitates the local search. The denotation $x_i(t)$ is the position of the particle and $v_i(t)$ is the velocity of the particle at $t$-th iteration, and $p_i(t)$ and $p_g(t)$ represent pbest and gbest. The $i$-th particle's position is denoted by $x_i(t) = (x_{i1}, x_{i2}, \ldots, x_{id})$ and the velocity is represented as $v_i(t) = (v_{i1}, v_{i2}, \ldots, v_{id})$ in a $d$-dimensional vector.

4.1. PSO algorithm for routing in our case:

Agent index is $a_i$ for an arbitrary $i$.
Particle index is $p_i$ for an arbitrary $i$.
**Step 1.** Initialize $a_i$ with the position and two velocities randomly.
**Step 2.** Find the fitness value of each $a_i$.
**Step 3.** Calculate pbest and gbest for each agent $a_i$.
**Step 4.** Do
Update the position and velocity of each particle:

$$v_i(t+1) = \omega v_i(t) + c_1 r_1 [p_i(t) - x_i(t)] + c_2 r_2 [p_g(t) - x_i(t)],$$
$$x_i(t+1) = x_i(t) + v_i(t+1).$$

Compute the fitness[$a_i$], the fitness value of each agent
If the current fitness value is better than the agent's pbest:
    update pbest of each agent $a_i$
    update gbest.
    best value:= gbest.
Repeat till the stop criterion.

At each iteration, the agent searches for the optimal solution by adjusting their properties.

The main drawback of PSO is that it easily drops into a local optima due to the fact that the particles rapidly get converged to the best particle. Many improvements and modifications have been introduced on the original PSO algorithm to avoid falling into the local optima.

## 5. Genetic algorithm

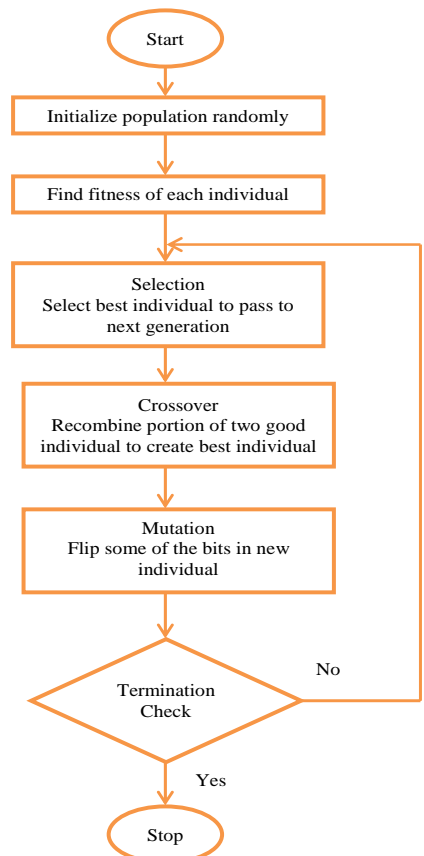

Fig. 1. Genetic algorithm

The Genetic algorithm [9] is a type of an evolutionary algorithm which uses operations, such as selection or reproduction, crossover or recombination, and mutation to produce solutions. The algorithm starts by generating initial population randomly and in each generation, the fitness function is evaluated for every individual; the individuals are picked up from the current generation based on the best fitness value and improved through crossover and mutation operations to create a new population for the next generation. The search process continues until a satisfactory fitness level or a finite number of generations is reached. The disadvantages of the Genetic algorithm are:
- There is always an issue in finding the fitness function.
- Designing the stopping criteria becomes a major issue.
- Not applicable for dynamic data, though the network today is more dynamic. Hence, this characteristic is a hurdle to use this algorithm in networking.
- Crossover and mutation operations are also difficult to define in networking.

## 6. Hybrid PSO-GA

The hybrid PSO-GA [10] integrates the strength of PSO with Genetic algorithms, and the hybrid algorithm combines the standard velocity and the position update rules of PSO with the ideas of selection and crossover from GAs. GA accomplishes a global search and PSO performs a local search. The Hybrid PSO-GA optimization algorithm searches the solution space optimally to reach the Gateway. The hybrid approach removes the weakness of PSO and GA, and also the balance of good knowledge sharing and natural selection to provide an efficient and optimal search in the solution space.

**Algorithm**

**Step 1.** Select the part of the best particles and keep it in a set called elitism.

**Step 2.** If $N$ is the total number of particles and $N_{\text{elitism}}$ is the particle in elitism set, select the particles using the following formula to apply PSO rules of the standard velocity and position update.

$$(3) \qquad (N - N_{\text{elitism}}) \times \text{Breed\_Ratio}.$$

Breed ratio is specified between 0 and 1.

**Step 3.** Apply the crossover and mutation operation on the remaining particles. The position of the particle is updated based on the Velocity Propelled Averaged Crossover (VPAC) method

$$(4) \qquad x'_p = \frac{x_p + x_q}{2 - \varphi_1 v_p}, \quad x'_q = \frac{x_p + x_q}{2 - \varphi_2 v_p},$$

where $\varphi_1$ and $\varphi_2$ can be taken within the range (0, 1); $x'_p$ and $x'_q$ are two children created by the particles $p$ and $q$; $x_p$ and $v_p$ are the current positions and velocities of the particle.

## 7. The mathematical model for QoS intelligent routing

In the communication graph, $G = (V, E)$, where $V$ depicts the set of routers and $E$ depicts the links between the routers, the edge $l_{ij}$ is between node $i$ and $j$. The QoS parameters are: the bandwidth $BW_{ij}$, the delay $D_{ij}$, the jitter $J_{ij}$ and the interference $I_{ij}$. There are many sources-destination pairs and many possible paths between the source and the destination. The routers are connected to the outside world through gateways and the gateway is responsible for sending and receiving the data. The QoS intelligent routing is used to find an optimal path from the source node to the gateway that reduces the cost and also satisfies the QoS constraints of bandwidth, delay, jitter and interference. The problem is to find a path from the source to the destination that reduces the cost subject to QoS constraints. The objective function $f(x)$ requires the least cost of the path $x$.

$$\text{(5)} \quad \text{Minimum } f(x) = \sum_{l_{ij} \in x} \cos t(l_{ij}),$$

s.t.

$$\text{(6)} \quad \min_{l_{ij} \in x} BW_{ij} \geq BW_{\text{req}},$$

$$\text{(7)} \quad \sum_{l_{ij} \in x} D_{ij} \leq D_{\text{req}},$$

$$\text{(8)} \quad \sum_{l_{ij} \in x} J_{ij} \leq J_{\text{req}},$$

$$\text{(9)} \quad I_{ij} \geq \beta.$$

Equation (6) gives the bandwidth constraints, (7) gives the delay constraints, (8) gives the jitter constraints and (9) gives the interference constraints.

The link in the graph has four weights BW, $D$, $J$, and $C$ which represent the bandwidth, delay, jitter and cost. The cost of the link is the sum of these four weights. $\beta$ is the interference threshold. The transmission on link1 with channel 1 can be viewed as interference to the transmission on link2 with an adjacent channel 2, and the interference is given by $I$-factor$(i, j)$ [11]. The Signal to Noise ratio is modelled as an $I$-factor.

### 7.1. Fitness functions

In QoS intelligent routing, the evolution is determined by the fitness function value which gives the quality of each particle. The fitness function is evaluated for each particle in individual swarms in each generation. The fitness function is determined by summing of the penalty and objective functions. The penalty function determines the degree of penalty for violating the QoS constraints. The penalty function $p(x)$ is determined as follows:

$$\text{(10)} \quad \begin{aligned} p(x) = &\eta_1 \max(BW_{\text{req}} - BW_{ij}, 0) + \eta_2 \max(D_{ij} - D_{\text{req}}, 0) + \\ &+ \eta_3 \max(J_{ij} - J_{\text{req}}, 0) + I\text{-factor}, \end{aligned}$$

where $\eta_1$, $\eta_2$ and $\eta_3$ are real numbers used for normalizing the bandwidth, delay and jitter, and these are called punishment coefficients. $BW_{\text{req}}$, $D_{\text{req}}$ and $J_{\text{req}}$ are the values of the bandwidth, delay and jitter specified by the application.

The fitness function for a hybrid PSO-GA is determined as follows:

(11) $$F(x) = f(x) + p(x).$$

If $p(x)$ value is 0, then the QoS constraints are satisfied and the packets are sent through the interference free path, otherwise $p(x)$ is between 0 and 1.

### 7.2. The Hybrid PSO-GA Routing

The input to PSO-GA algorithm is specified in the procedure of the particle. The multiple routes between the source and the gateway are encoded as a particle, i.e., the sequence of nodes is represented as a particle which is encoded as an integer value. So the subtraction between two positions in (1) and the addition of a position and velocity in (2) are not suitable for this problem. The particles are initialized with a random position and velocity. The breed ratio determines the amount of population, which undergoes PSO or GA. The value of the breed ratio ranges from 0.0 up to 1.0. The breed ratio is set to 0.5, so the half of the particle is updated by PSO, and the remaining half is updated by GA simultaneously. In Fig. 2 the flowchart of the hybrid algorithm is represented.

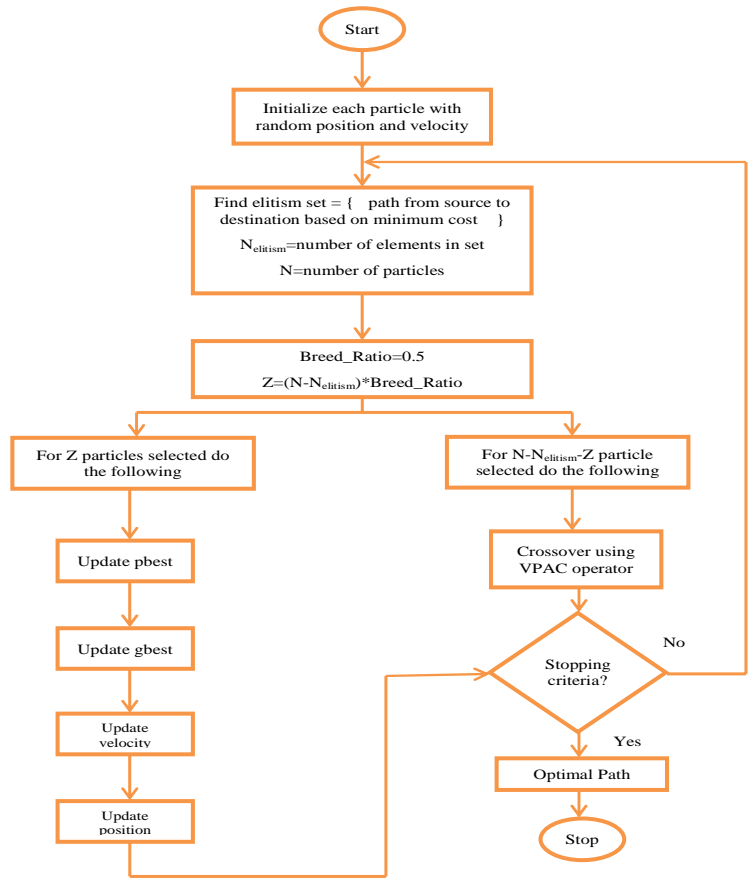

Fig. 2. The Hybrid PSO-GA routing algorithm

**Algorithm**

**Step 1.** Initialize each particle with a random position and velocity.

**Step 2.** Find an initial solution from the source to the destination based on the minimum cost from one node to another and put all the nodes accessed into a set called elitism.

**Step 3.** If $N$ is the total number of nodes and $N_{elitism}$ is the number of nodes in the elitism set then select the following number of nodes from the set other than the elitism set

(12) $$Z = (N - N_{elitism}) \times \text{Breed\_Ratio}.$$

**Step 4.** If $Z$ has a decimal value around it, select $Z$ nodes randomly and update the velocity and position as follows:

(13) $$x_i(t+1) = x_i(t) \oplus c_1 r_1(p_i(t)) \oplus c_2 r_2(p_g(t)),$$

$x_i(t)$ is the sequence of nodes expressed by a particle, $p_i(t)$ is pbest and $p_g(t)$ is gbest.

**Step 5.** Now the left nodes other than the elitism set and PSO updating, the rest of the nodes are updated using GA crossover operations.

7.2.1. $\oplus$ Operator

Assume that $p_i = (x_1, x_2, x_3, \ldots, x_k)$ and $p_g = (y_1, y_2, y_3, \ldots, y_k)$.
For example if $p_i = (1, 2, 4, 9, 13)$ and $g_i = (1, 7, 5, 10, 13)$
$$P_a' = p_i \oplus p_g,$$
$$P_a' = \{1, \text{alter}(2, 7), \text{alter}(4, 5), \text{alter}(9, 10), 13\},$$
$$P_a' = \{1, 7, 5, 9, 13\},$$
Alter$(2, 7) = \min \{(s, 2), (s, 7)\}$, where $s$ is the source node. Find the minimum cost of the source to node 2 and 7. Node 7 is having the minimum cost path from the source node, so that node 7 is included in the set and node 2 is eliminated. Repeated nodes in $P'_a$ are eliminated.

7.2.2. Crossover operator

Two particles are selected randomly from the population for a two point crossover. Two points are selected for the crossover, a sequence of nodes from the beginning of the particle to the first crossover point is selected, the part of the particle from the first point to the second point is selected from the second particle, and the remaining is copied from the first particle.

For example
$$P_1 = \{1, 7, 5, 8, 12, 15, 21, 24, 25\};$$
$$P_2 = \{1, 7, 5, 10, 17, 19, 22, 25\};$$
$$P'_1 = \{1, 7, 5, 10, 12, 15, 21, 24, 25\};$$
$$P'_2 = \{1, 7, 5, 8, 12, 15, 21, 25\}.$$

Sometimes the crossover operator and $\oplus$ operator lead to unconnected route, so that we need to be careful while finding the fitness values of this particle. At each iteration there exist two or more redundant particles. These duplicate particles are discarded at each iteration to increase the searching ability.

## 8. Simulation

Java and JADE framework is used to simulate the QoS intelligent routing algorithm. The network topology taken for the optimal routing is shown in Fig. 3, comprising of 25 nodes. The node 1 is the source node and there are 3 Gateways to connect to Internet, which are nodes 11, 13 and 25. The performance of the intelligent routing algorithm is tested for 50, 75, 100 and 125 nodes. Each node is equipped with multiple network interfaces which are tuned to multiple channels. Many possible routes are available between the source and the gateway when the network size is larger or it is densely connected.

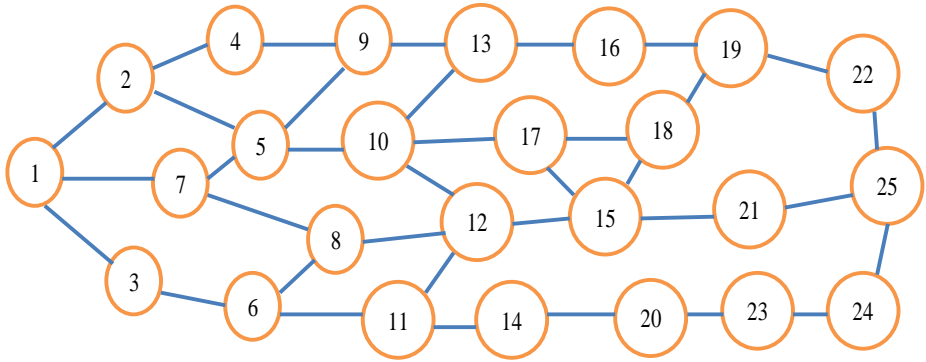

Fig. 3. Random topology

The physical distance between any two nodes differs randomly. The transmission range of the nodes is set to 250 meters. A link exists between the two nodes if it is within the hearing range of each other. The link cost is specified in the cost matrix within the range of [2-10] and the bandwidth is set to 11 Mbps uniformly for all the links. Similarly, the delay matrix within the range of [0.5-2 ms], the packet loss matrix within the range of [0.001-010] and the jitter matrix is in the range of [0.5-2.0 ms].

The interference value is normalized between 0 and 1, and assigned to each link which is specified in *I*-matrix. The $BW_{req}$, $D_{req}$, $J_{req}$ values differ from application to application. The different source and destination nodes are selected for various runs in the same test.

### 8.1. Simulation results

Fig. 4 depicts that the value of fitness vs number of iterations for 25 nodes. The performance of PSO, GA, Hybrid is evaluated at the 14th iteration; Hybrid PSO-GA gives an optimal path whose fitness is 12.56. Fig. 4 shows the progress of the algorithm finding the optimal path for topology given in Fig. 3. PSO finds the optimal path at 20th iteration, but GA gets 16.56 at this iteration, which is a global optimal route, so GA needs some more time to converge. Table 1 indicates the path taken and the fitness value at each iteration.

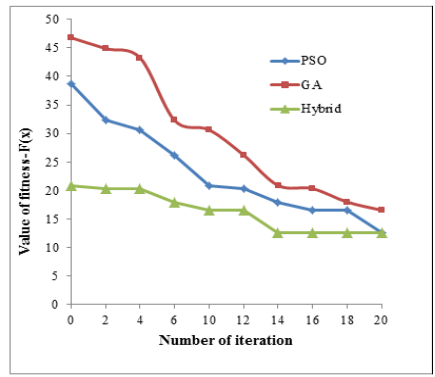

Fig. 4. The value of fitness vs number of iterations

Fig. 5 shows the computation time for PSO, GA, Hybrid with increase of the number of nodes in the network. It indicates that the computation speed of all three algorithms reduces, when more nodes are added to the network. The hybrid algorithm shows better performance compared to PSO and GA, it yields the optimal solution quickly, when more than 100 nodes are added to the network.

Table 1. Path taken at each iteration

| Iteration | Path | Fitness value |
|---|---|---|
| 1 | 1-7-5-10-13 | 20.9 |
| 2-4 | 1-7-8-12-11 | 20.34 |
| 6-10 | 1-3-6-11 | 18 |
| 10 | 1-2-4-9-13 | 16.56 |
| 11-13 | 1-2-4-9-13 | 16.56 |
| 14-20 | 1-7-5-9-13 | 12.56 |

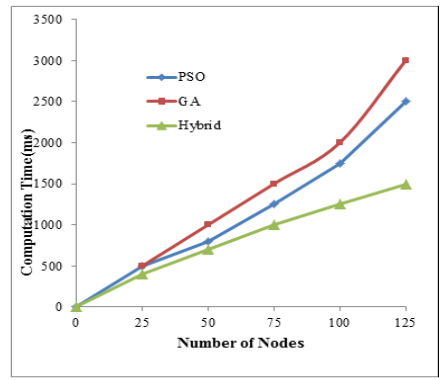

Fig. 5. The convergence time vs number of nodes

Fig. 6 shows the packet delivery ratio vs number of nodes. The Hybrid gives better performance compared to PSO and GA, the packet delivery ratio is retained at 90% even after the network size reaches 100 nodes. It shows that the hybrid approach guarantees QoS and is more suitable for reliable communication in MCMR-WMN.

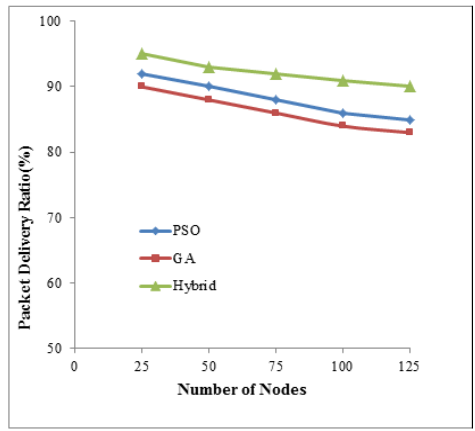

Fig. 6. The packet delivery ratio vs number of nodes

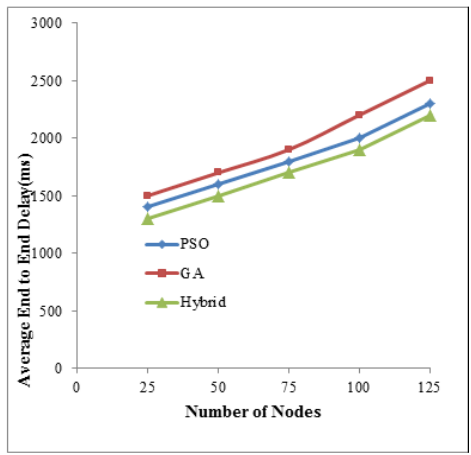

Fig. 7. The average end-to-end delay vs number of nodes

Fig. 7 shows the average end-to-end delay vs number of nodes. The delay increases gradually when the number of nodes in the network increases. The hybrid approach gives a smaller average end-to-end delay compared to PSO and GA.

The Hybrid algorithm outperforms the PSO and GA in terms of the convergence time, packet delivery ratio and the average end-to-end delay. Thus, it is evident that the Hybrid PSO-GA is very suitable for optimization of the routing in MCMR-WMN with POC assigned. The other techniques like PSO and GA fail to find the optimal solution in a large WMN with multiple constraints.

## 9. Conclusion

The QoS guarantee is essential for real time communications, but it is hard to achieve QoS in Wireless Networks. In this paper an intelligent routing, using the hybrid PSO-GA is proposed to support QoS. The Hybrid algorithm removes the weakness of PSO and GA, and it increases the stability between the knowledge sharing and the natural selection to find the optimal solution in the search space. Half of the particle is updated by a standard position and velocity update of PSO and the remaining half is updated by a crossover operation of GA simultaneously. The QoS parameter and the interference value are added into the fitness function to find the optimal path. The simulation results show that the Hybrid algorithm efficiently solves QoS routing, also giving smaller convergence time, end-to-end delay and better delivery ratio.